\documentclass[english,pra,showpacs,showkeys,tightenlines,secnumarabic,11pt]{revtex4}
\usepackage[T1]{fontenc}
\usepackage[latin1]{inputenc}
\usepackage{amsmath}
\usepackage{graphicx}
\usepackage{amssymb}
\usepackage{epsfig}
\usepackage{graphics}
\usepackage[mathscr]{euscript}
\usepackage{psfrag}
\usepackage{pstricks}
\usepackage{pst-node}
\usepackage{babel}

\begin{document}
\title{Fractional momentum correlations in multiple production of
W bosons and of ${b\bar b}$ pairs in high energy $pp$ collisions}
\author{E. Cattaruzza}
\email{ecattar@ts.infn.it}
\author{A. Del Fabbro}
\email{delfabbr@ts.infn.it}
\author{D. Treleani}
\email{daniel@ts.infn.it}
\affiliation{ Dipartimento di Fisica Teorica dell'Universit\`a di Trieste and
INFN, Sezione di Trieste,\\ Strada Costiera 11, Miramare-Grignano,
I-34014 Trieste, Italy.}
\begin{abstract}
Multiple parton collisions will represent a rather common feature
in $pp$ collisions at the LHC, where regimes with very large
momentum transfer may be studied and events rare in lower energy
accelerators might occur with a significant rate. A reason of
interest in large $p_t$ regimes is that, differently from low
$p_t$, evolution will induce correlations in $x$ in the
multiparton structure functions. We have estimated the cross
section of multiple production of $W$ bosons with equal sign,
where the correlations in $x$ induced by evolution are
particularly relevant, and the cross section of $b\bar b b\bar b$
production, where the effects of evolution are much smaller. Our
result is that, in the case of multiple production of $W$ bosons,
the terms with correlations may represent a correction of the
order of 40\% of the cross sections, for $pp$ collisions at 1 TeV
c.m. energy, and a correction of the order of 20\% at 14 TeV. In
the case of $b\bar b$ pairs the correction terms are of the order
of $10-15\%$ at 1 TeV and of the order of 5\% at 14 TeV.
\end{abstract}
 \maketitle

\section{Introduction}
Multiple parton interactions in high energy hadronic collisions
have been discussed long ago by several
authors \cite{Landshoff,Paver}. Experimentally events with
multiparton interactions have been first observed in $pp$
collisions by the AFS Collaboration  \cite{Akesson:1986iv} and
later, with sizably larger statistics, at Fermilab by the CDF
Collaboration \cite{CDF}.

In multiple parton collisions the hadron is probed in different
points contemporarily \cite{Paver}. The non trivial feature of
multiple parton collisions is hence its non-perturbative input,
which has a direct relation with the correlations between partons
in the hadron structure \cite{Calucci:1997ii}. As the process is
originated by the large population of partons in the initial
state, the expectation is nevertheless that correlations should
not represent a major feature in the process, with the exception
of correlations in transverse space, which are directly measured
by the cross section. Indeed the experimental analysis and most of
the theoretical estimates have been done with this simplifying
assumption and, although the statistics was too low to draw firm
conclusions, the CDF Collaboration reported that the cross section
is not influenced appreciably when changing the fractional momenta
of initial state partons \cite{CDF}.

The much larger rates of multiple parton collisions expected at
the LHC, with the possibility of testing different multiparton
processes at different resolution scales, represents however a
good motivation for reconsidering the approach to the problem. In
particular an interesting process, where multiple parton
collisions play an important role and which might be observed at
the LHC, is the production of multiple $W$ bosons with equal
sign \cite{Stirling1}, which would allow testing multiple parton
interactions at a much larger resolution scale than usually
considered. The evolution of the multiparton structure functions
will play a non minor role in this case, leading to sizable
correlations in fractional momenta.

The purpose of the present note is to give some quantitative
indication of the effects in multiparton collisions in a high
resolution regime and compare with a case at a lower resolution.
After recalling the basic features of the inclusive cross section
of double parton collisions, we will hence evolve double parton
distributions at high resolution scales. The effect of
correlations induced by evolution will be estimated studying the
cross sections of multiple production of equal sign $W$ bosons and
the cross section of multiple production of $b \bar b$ pairs, in
the energy range $1-14$ TeV.

\section{Double parton cross section}

With the only assumption of factorization of the two hard parton
processes A and B, the inclusive cross section of a double
parton-scattering process in a hadronic collision is expressed
by \cite{Paver,Braun}
\begin{equation}
    \sigma^D_{(A,B)}=   \frac{m}{2}\sum_{i,j,k,l}\int\Gamma_{ij}(x_1,x_2,b)\,\hat{\sigma}^A_{ik}(x_1,x_1')
    \,\hat{\sigma}^B_{jl}(x_2,x_2')\, \Gamma_{kl}(x_1',x_2';b)\,dx_1\,dx_1'\,dx_2\,dx_2'\,d^2b,
    \label{sigmaD}
\end{equation}
where $\Gamma_{ij}(x_1,x_2,b)$ are the double parton distribution functions, depending on
the fractional momenta $x_1,\,x_2$ and on the relative transverse distance $b$ of the two
partons undergoing the hard processes A and B, the indices $i$ and $j$ refer to the
different parton species and $\hat{\sigma}^A_{ik}$ and $\hat{\sigma}^B_{jl}$ are the
partonic cross sections. The dependence on the resolution scales is implicit in all
quantities. The factor $m/2$ is a consequence of the symmetry of the expression for
interchanging $i$ and $j$; specifically $m=1$ for indistinguishable parton processes and
$m=2$ for distinguishable parton processes.

The double distributions $\Gamma_{ij}(x_1,x_2,b)$ are the main
reason of interest in multiparton collisions. The distributions
$\Gamma_{ij}(x_1,x_2,b)$ contain in fact all the information of
probing the hadron in two different points contemporarily, though
the hard processes A and B.

The cross section for multiparton process is sizable when the flux
of partons is large, namely at small $x$, and dies out quickly for
larger values. Given the large parton flux one may hence expect
that correlations in momentum fraction will not be a major effect
and partons to be rather correlated in transverse space (as they
must anyhow all belong to the same hadron). Neglecting the effect
of parton correlations in $x$ one writes
\begin{equation}
    \Gamma_{ij}(x_1,x_2;b)=\Gamma_i(x_1)\,\Gamma_j(x_2)\,F^i_j(b),
    \label{factansatz}
\end{equation}
where $\Gamma_i(x)$ are the usual one body parton distribution function and $F^i_j(b)$ is
a function normalized to one and representing the parton pair density in transverse
space. The inclusive cross section hence simplifies to
\begin{equation}
    \sigma^D_{(A,B)}=\frac{m}{2}\sum_{ijkl} \Theta^{ij}_{kl}\,\hat{\sigma}_{ij}(A)\,\hat{\sigma}_{kl}(B),
\end{equation}
where $\hat{\sigma}_{ij}(A)$ and $\hat{\sigma}_{kl}(B)$ are the
hadronic inclusive cross sections for the two partons labelled $i$
and $j$ to undergo the hard interaction labelled $A$ and for two
partons $k$ and $l$ to undergo the hard interaction labelled $B$;
\begin{equation}
    \Theta^{ij}_{kl}=\int d^2b\,F^i_k(b)\,F^j_l(b)
\end{equation}
are geometrical coefficients with dimension an inverse cross
section and depending on the various parton processes. In the
simplified scheme above, the coefficients $\Theta^{ij}_{kl}$ are
the experimentally accessible quantities carrying the information
of the parton correlations in transverse space.

In the experimental search of multiple parton collisions the cross
section has been further simplified assuming that the densities
$F^i_j$ do not depend on the indices $i$ and $j$, which leads to
the expression
\begin{equation}
    \sigma^D_{(A,B)}=\frac{m}{2}\,\frac{\hat{\sigma}(A)\,\hat{\sigma}(B)}{\sigma_{eff}}\equiv \sigma_{fact}^D,
\label{fact-equation}
\end{equation}
where all information on the structure of the hadron in transverse
space is summarized in the value of a single the scale factor,
$\sigma_{eff}$. In the experimental study of double parton
collisions CDF quotes $\sigma_{eff}=14.5\,mb$  \cite{CDF}.

The experimental evidence is not inconsistent with the simplest
hypothesis of neglecting correlations in momentum fractions, the
resolution scale probed in the CDF experiment is however not very
large, the transverse momenta of final state partons being of the
order of 5 GeV. We will hence approach the problem in more general
terms, focusing on multiple production of equal sign $W$ bosons
and of $b\bar b b\bar b$ pairs, keeping into account the
correlations in fractional momenta induced by evolution.

\section{Two-body distribution functions}

The evolution of the double parton distribution function has been discussed in
refs \cite{Kirschner,Shelest} and more recently in \cite{Snigirev}. The approach is
essentially the same used to study particle correlations in the fragmentation
functions \cite{Puhala}, using the jet calculus rules \cite{Konishi}.

Introducing the dimensionless variable
\begin{equation}
    t=\frac{1}{2\,\pi\,b}\ln \left[1+\frac{g^2(\mu^2)}{4\,\pi}b\,\ln\left(\frac{Q^2}{\mu^2}\right)\right],\,\,\,b=\frac{33-2\,n_f}{12\,\pi},\nonumber
\end{equation}
where $g^2(\mu^2)$ is the running coupling constant at the
reference scale $\mu^2$ and $n_f$ the number of active flavors,
the probability $D_h^{j_1j_2}(x_1,x_2;t)$  to find two partons of
types $j_1$ and $j_2$ with fractional momenta $x_1$ and $x_2$
satisfy the generalized Lipatov-Altarelli-Parisi-Dokshitzer
evolution equation
\begin{align}
    \frac{dD_h^{j_1j_2}(x_1,x_2;t)}{dt}& =\sum_{j_1'}\int_{x_1}^{1-x_2}\,\frac{dx_1'}{x_1'}\,D_h^{j_1'j_2}(x_1',x_2;t)
    \,P_{j_1'\to j_1}\left(\frac{x_1}{x_1'}\right)\nonumber\\
     & +\sum_{j_2'}\int_{x_2}^{1-x_1}\,\frac{dx_2'}{x_2'}\,D_h^{j_1j_2'}(x_1,x_2';t)
    \,P_{j_2'\to j_2}\left(\frac{x_2}{x_2'}\right)\nonumber\\
     & +\sum_{j'}D_h^{j'}(x_1+x_2;t)\frac{1}{x_1+x_2}\,P_{j'\to j_1j_2}\left(\frac{x_1}{x_1+x_2}\right),
    \label{L-A-P-D}
\end{align}
where the subtraction terms are included in the evolution kernels $P$.

If at the scale $\mu^2$ one assumes the factorized form
\begin{equation}
D_h^{j_1j_2}(z_1,z_2,0)=D_h^{j_1}(z_1;0)\,D_h^{j_2}(z_2;0)\,\theta(1-z_1-z_2),
\end{equation}
at a larger scale one obtains a solution which may be expressed as
the sum of a factorized and of two non-factorized contributions:
\begin{equation}
    D_h^{j_1j_2}(x_1,x_2;t)=D_h^{j_1}(x_1;t)\,D_h^{j_2}(x_2;t)\,\theta(1-x_1-x_2)
    +D_{h,corr,1}^{j_1j_2}(x_1,x_2;t)+D_{h,corr,2}^{j_1j_2}(x_1,x_2;t),
    \label{solution}
\end{equation}
where the non-factorized contributions are expressed by the convolutions:
\begin{align}
    D_{h,corr,1}^{j_1j_2}(x_1,x_2;t)&=\theta(1-x_1-x_2)\,
    \left[\sum_{j_1'j_2'}\int_{x_1}^{1}\frac{dz_1}{z_1}
    \int_{x_2}^1\frac{dz_2}{z_2}\,D_h^{j_1'}(z_1,0)\,D_{j_1'}^{j_1}(\frac{x_1}{z_1};t)\right.\nonumber\\
    &\times \left.
    D_h^{j_2'}(z_2,0)\,
    D_{j_2'}^{j_2}(\frac{x_2}{z_2};t)\,\,[\theta(1-z_1-z_2)-1]\right]
    \label{correlation1}\\
    D_{h,corr,2}^{j_1j_2}(x_1,x_2;t)&=\sum_{j'j_1'j_2'}\int_0^tdt'\int_{x_1}^{1}\frac{dz_1}{z_1}
    \int_{x_2}^{1-x_1}\frac{dz_2}{z_2}\,D_h^{j'}(z_1+z_2;t')
    \frac{1}{z_1+z_2}\nonumber\\
    &\times\,P_{j'\to j_1'j_2'}
    \left(\frac{z_1}{z_1+z_2}\right)
    \,D_{j_1'}^{j_1}(\frac{x_1}{z_1};t-t')\,
    D_{j_2'}^{j_2}(\frac{x_2}{z_2};t-t');
    \label{correlation2}
\end{align}
and the distribution functions $D_i^j(x;t)$ satisfy the evolution
equation
\begin{equation}
\frac{dD_i^j(x;t)}{dt}=\sum_{j'}\int_x^1\frac{dx'}{x'}D_i^{j'}(x';t)\,P_{j'-j}\left(\frac{x}{x'}\right).
\label{Altarelli}
\end{equation}
with initial condition $D_i^j(x;t=0)=\delta_{ij}\,\delta(1-x)$

Equations(\ref{Altarelli}) are solved by introducing the Mellin transforms
\begin{equation}
D_i^j(n;t)=\int_0^1 dx\,x^n\,D_i^j(x;t),
\end{equation}
which lead to a system of ordinary linear-differential equations
at the first order. The solution is given by the inverse Mellin
transform
\begin{eqnarray}
D_i^j(x;t)&=&\int \frac{dn}{2\,\pi\,\imath} x^{-n}\,D_i^j(n;t)=\mathcal{L}^{-1}(D_{i}^j(n;t),-\ln(x)),
\end{eqnarray}
where the integration runs along the imaginary axis at the right
of all the $n$ singularities, while $\mathcal{L}^{-1}$ represents
the Inverse Laplace operator.

The double distributions can then be obtained numerically. For
inverting the Laplace Transform we have followed two different
procedures \cite{Abate}: the Gaver-Wynn-Rho (GWR) algorithm and the
fixed Talbot (FT) method. The first procedure (GWR) is based on a
special acceleration sequence of the Gaver functionals and
requires to evaluate the transform only on the real axes; the
second procedure (FT) is based on the deformation of the contour
of the Bromwich inversion integral and requires complex
arithmetic. Comparing the two methods we have found more stable
results when using the (FT) method. The double distributions have
hence been obtained by numerical integration with the Vegas
algorithm  \cite{Vegas}, using the MRS99  \cite{MRS99} as input
parton distribution function at the scale $\mu^2$.

In the kinematical range of interest for the actual case (we never exceed $x=.1$) the
contribution of the term $D_{h,corr,1}^{j_1j_2}$ in eq.(\ref{solution}) is negligible.
The first term in eq.(\ref{solution}) represents the factorized contribution usually
considered and is the solution of the homogeneous (LAPD) evolution equation, while the
third term is a particular solution of the complete equation.

\begin{figure}[t]
\begin{center}
\includegraphics[scale=0.5,angle=270]{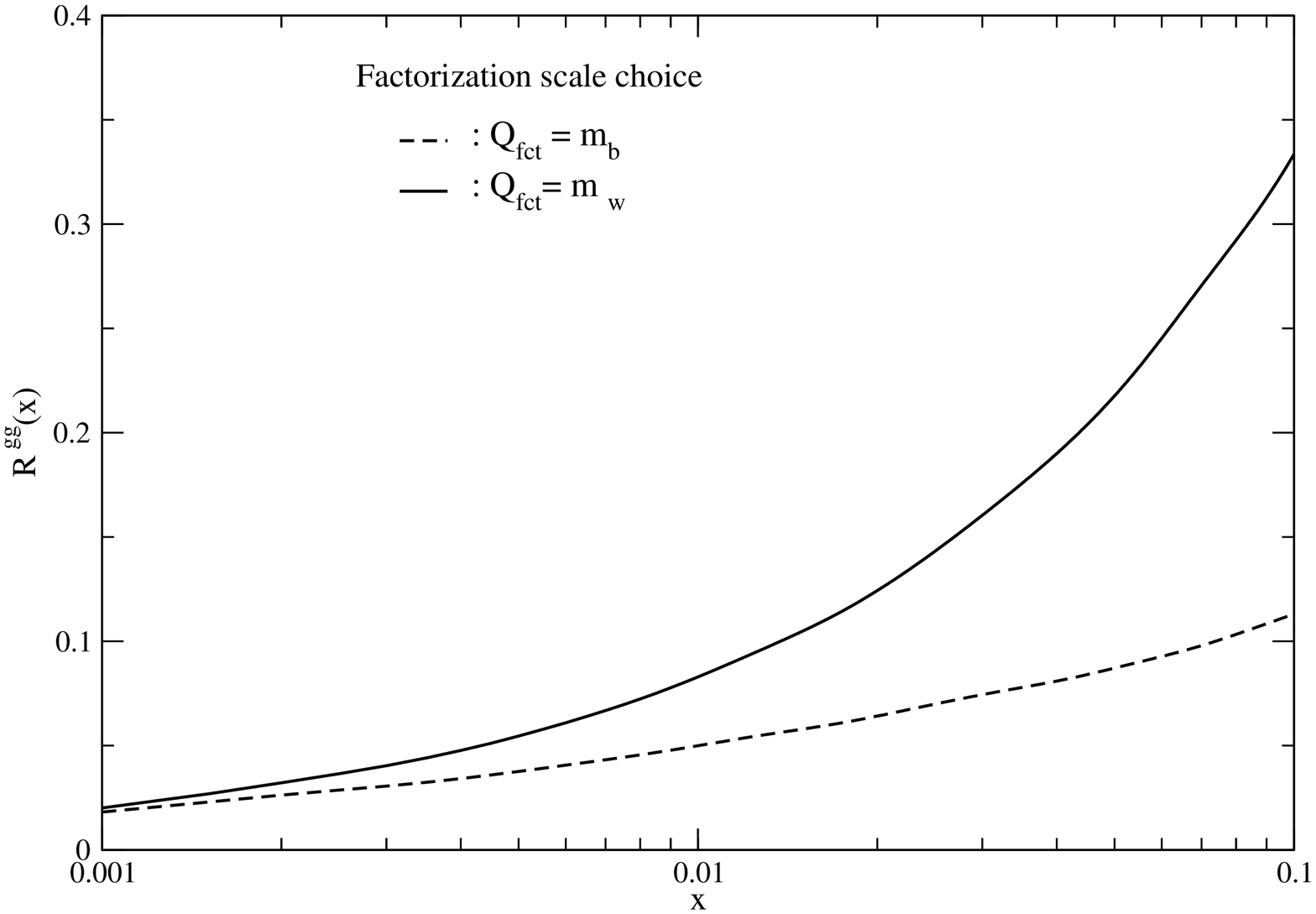}
\caption{ $gg$ correlation ratio for $x_1=x_2=x$, with factorization scale equal to the
$W$ mass (solid- curve) and to the $b$ mass (dashed-curve).}
\label{gg-correlation}
\end{center}
\end{figure}

\begin{figure}[t]
\begin{center}
\vskip0.5cm
\includegraphics[scale=0.5,angle=270]{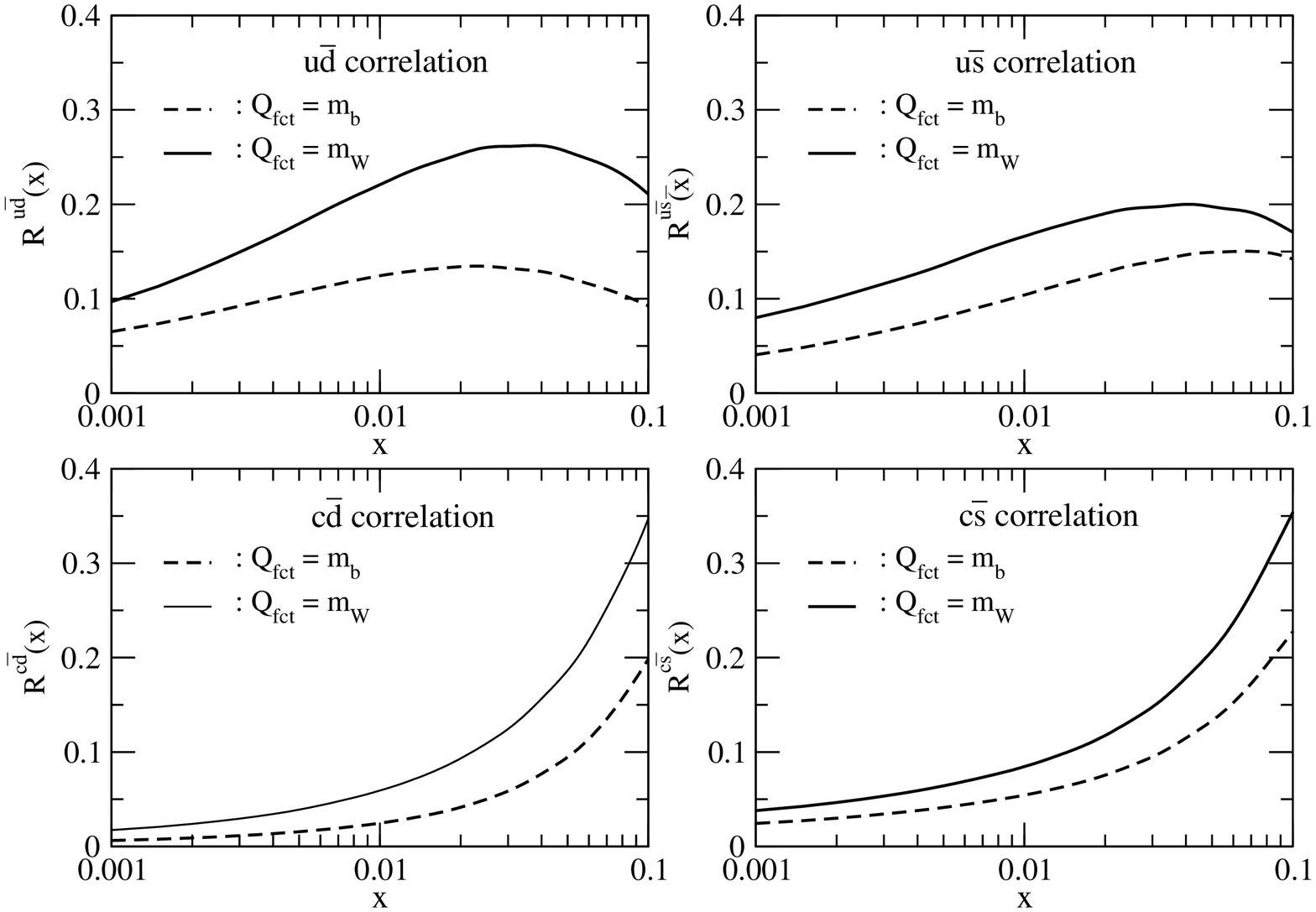}
\caption{ $qq$ correlation ratios for $x_1=x_2=x$, with
factorization scale equal to the $W$ mass (solid- curves) and to
the $b$ mass (dashed-curves).} \label{qq-correlation}
\end{center}
\end{figure}

The effect of the correlation terms induced by evolution is shown for gluon-gluon and for
quark-quark in Fig.[\ref{gg-correlation},\ref{qq-correlation}], where the ratio
\begin{equation}
R^{j_1j_2}(x_1,x_2;t)=\frac{D_{h,corr,1}^{j_1j_2}(x_1,x_2;t)+D_{h,corr,2}^{j_1j_2}(x_1,x_2;t)
}{D_h^{j_1}(x_1;t)\,D_h^{j_2}(x_2;t)}.
\end{equation}
is plotted as a function of $x$, with $x_1=x_2=x$, with the
following choice of parameters: $\mu=1.2\,GeV$, $n_f=4$,
factorization scale equal to the $W$ mass, $m_w=80.4\,GeV$ (solid
curves) and factorization scale equal to the bottom quark mass,
$m_b=4.6\,GeV$ (dashed curves).

As shown in Fig.[\ref{gg-correlation}], the ratio $R^{gg}$ is
nearly $35\%$ for $x\sim 0.1$ and decreases up to $8-10\%$ for
$x\sim 0.01$ and to $2\%$ for $x\sim 0.001$, when the $W$ mass is
used as factorization scale. When taking the $b$ quark mass as
factorization scale, the value of the ratio is of the order of
$10-12\%$ for $x\sim 0.1$ and decreases up to $5\%$ and to $2\%$
for $x\sim 0.01$ and $x\sim 0.001$ respectively. The ratio would
of course be much larger (up to $60\%$) if going to larger $x$
values.

The ratio $R^{qq}$ is shown in Fig.[\ref{qq-correlation}] for a
few flavor choices. With the $W$ mass as factorization scale, the
ratios are of the order of $35,\,20,\,10\%$ for $x\sim
0.1,\,0.01,\,0.001$. With the $b$ quark mass as factorization
scale the ratios are of order of $23,\,10,\,5\%$ for $x\sim
0.1,\,0.01,\,0.001$.\\
Apart from the case of hadron-nucleus collisions, when two
different target nucleons take part to the
process \cite{Strikman:2001gz}, the non-perturbative input of the
double parton scattering cross section is not represented however
by the distribution functions $D^{j_1j_2}_h(x_1,x_2;t)$ in
Eq.(\ref{solution}), where all transverse variables have been
integrated. The double parton scattering cross section,
Eq.(\ref{sigmaD}), depends in fact in a direct way also on the
relative separation of partons in transverse space, which is of
the order of the hadron size and hence outside the control of
perturbation theory.

Considering that the longitudinal and the transverse momenta of
initial state partons are essentially decoupled in the process,
because of the different scales involved, it's not unreasonable to
assume phenomenologically a factorized dependence of the double
distribution functions on the longitudinal and transverse degrees
of freedom. Given the different origin of the terms in
$D_h^{j_1j_2}$, it's also not unnatural to consider the
possibility of having different non-perturbative scales, for the
transverse separation of the factorized and of the correlated
terms. In fact, although in the general case evolution would mix
the two scales in the $D_{h,corr,1}^{j_1j_2}$ term, the term
$D_{h,corr,1}^{j_1j_2}$ is very small in the kinematical regime of
interest and the hypothesis of two different transverse scales is
not inconsistent.

We hence assume that the typical transverse distance between partons in
$D_{h,fact}^{j_1j_2}$ and in $D_{h,corr,1}^{j_1j_2}$ corresponds to the relatively low
resolution scale process observed by CDF and, to have an idea on the effects of the
presence of two different scales in the double parton densities, we introduce a different
transverse distance in the term $D_{h,corr,2}^{j_1j_2}$, related to the size of the gluon
cloud of a valence quark, and corresponding to a relatively shorter range correlation
term. The double parton distributions are hence expressed in the following way:
\begin{equation}
D_h^{j_1j_2}(x_1,x_2;b;t)=
\left(D_{h,fact}^{j_1j_2}(x_1,x_2;t)+D_{h,corr,1}^{j_1j_2}(x_1,x_2;t)\right)\,F_{\sigma_{eff}}(b)
+D_{h,corr,2}^{j_1j_2}(x_1,x_2;t)\,F_{\sigma_{r}}(b)\nonumber
\end{equation}
where the parton pair densities $F_i(b)$ satisfy
\begin{equation}
\int d^2b \,F_i(b) = 1 \quad \int d^2b \,F_i(b)^2 = \frac{1}{\sigma_i}\nonumber
\end{equation}
with $i=\sigma_{eff},\,\sigma_{r}$.

While $F_{\sigma_{eff}}$ represents the transverse density of
partons at a relatively low resolution scale, relevant in the
kinematical conditions of the CDF experiment and leading to the
measured value of the scale factor $\sigma_{eff}=14.5$ mb,
$F_{\sigma_{r}}$ is rather the transverse parton density
characterizing partons correlated in fractional momentum, which
becomes increasingly important when the resolution scale is large.
To study the effect of the two scales we have let the smaller
scale vary in the interval $\sigma_{r_0} \leftrightarrow
\sigma_{eff}$ assuming $\sigma_{r_0}=2.8$ mb \cite{Povh}, which
might represent the size of the gluon cloud of a valence quark in
the hadron. To disentangle the effects of the correlation in
fractional momenta we have neglected a possible dependence of the
parton pair densities $F_i(b)$ on the partons flavor.

\section{Multiple production of $b\bar b$ pairs and of equal sign $W$ bosons in $pp$ collisions}

For the purpose of the present analysis we have hence evaluated
the contributions to multiple production of equal sign $W$ bosons
and to multiple production of $b\bar b$ pairs, due to multiple
(disconnected) parton collision processes, taking into account the
correlation terms in fractional momenta induced by evolution.

As a matter of fact higher order corrections in $\alpha_S$ are
very important in heavy quark production. To the purpose of the
present analysis we have evaluated the cross section at the lowest
order in perturbation theory, taking higher order corrections into
account by rescaling the lowest order results with a constant
factor $K$, defined as the ratio between the inclusive
cross-section for $b\bar b$ production, $\sigma(b\bar b)$, and the
result of the lowest-order calculation in pQCD. Our assumption is
hence that higher order corrections in $b\bar b b\bar b$
production may be taken into account by multiplying the cross
section of each connected process by the same factor $K$, so that
higher order corrections are taken into account by multiplying the
lowest order cross section by the $K$-factor at the second power.
In the actual calculation we have used a $K$ factor equal to
$5.7$ (\cite{Cattaruzza,DelFabbro4}) and the value $m_b=4.6\,GeV$
for the mass of the bottom quark.

The multiparton distributions have been obtained, as described in
the previous paragraph, using as input distributions at the scale
$\mu^2$ the MRS99 \cite{MRS99} parton distribution functions.
Factorization and renormalization scale have been set equal to the
transverse mass of the produced quarks. As for the dependence on
the transverse variables, in addition to the usual factorized
contribution, leading to the scale factor $1/\sigma_{eff}$, in the
present case the cross section includes also non factorized
contributions, corresponding to the couplings of
$D_{h,corr,2}^{ik}$ both with $D_{h,fact,1}^{jl}$ and with
$D_{h,corr,2}^{ik}$. We have assumed a gaussian distribution for
$F_{\sigma_{eff}}(b)$ and for $F_{\sigma_{r}}(b)$. The scale
factors are correspondingly $2/(\sigma_{eff}+\sigma_r)$ and
$1/\sigma_r$.

\begin{figure}[b]
\begin{center}
\includegraphics[scale=0.55,angle=270]{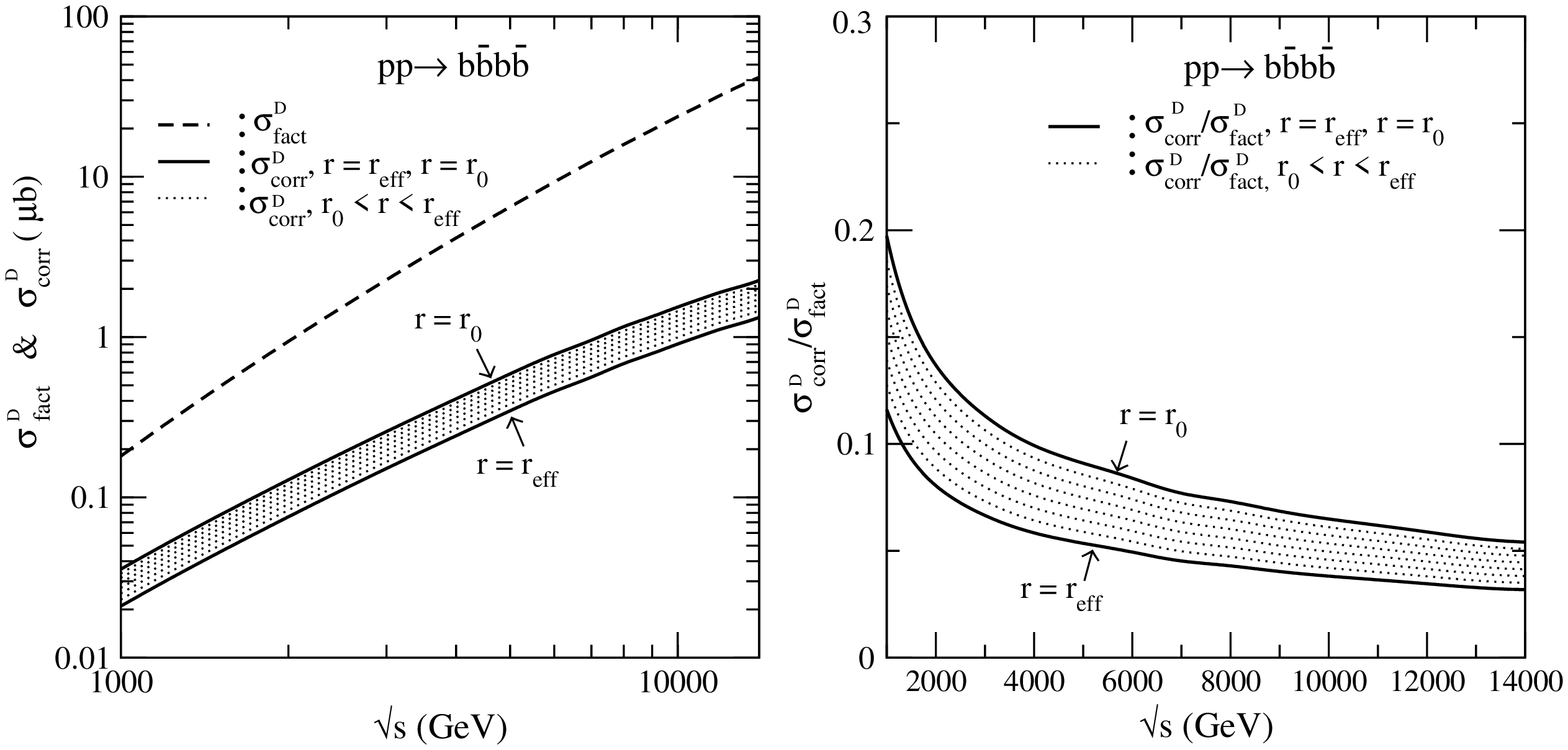}
\caption{\emph{Left-panel}: $b\bar bb\bar b$ inclusive cross-section as a function of
center of mass energy; the dashed curve refers to the double-parton scattering factorized
term $\sigma_{fact}^D$, while the upper and the lower solid curves refer to contributions
with correlations $\sigma_{corr}^D$, with geometrical factors obtained setting $r=r_0$
and $r=r_{eff}$ respectively. \\ \emph{Right-panel}: ratio between $\sigma_{corr}^D$ and
$\sigma_{fact}^D$ as a function of center of mass energy.}
\label{bb-spectra}
\end{center}
\end{figure}

In Fig.[\ref{gg-correlation}] we plot the $gg$ correlation (the
dominant contribution to $b\bar b$ is gluon fusion) while the
expected rise of the total $b\bar b b\bar b$ cross-section is
plotted in Fig.[\ref{bb-spectra}] (\emph{left-panel}) as a
function of the center of mass energy. The dashed curve refers to
the double-parton scattering factorized term ($\sigma_{fact}^D$)
given by eq.(\ref{fact-equation}); the continuous curves refer to
the double-parton scattering correlation contributions
($\sigma_{corr}^D$), with geometrical factors determined  by
setting $r=r_0$ (upper curve) and $r=r_{eff}$ (lower curve). The
ratio between the contribution of the terms with correlations and
the factorized term is shown in Fig.[\ref{bb-spectra}]
(\emph{right panel}) as a function of center of mass energy. The
effect of the terms with correlations decreases by increasing the
center of mass energy; depending on the values of
$r\in[r_{eff},r_0]$, correction effects may vary between
$(12-20)\%$ at $\sqrt s=1\,TeV$ and $(3.5-6)\%$ at $\sqrt
s=14\,TeV$. The decrease is faster as $\sqrt s\le 5\,TeV$: for
larger c.m. energies the average fractional momentum $<x>$ becomes
smaller than $0.01$, where the fraction $R^{gg}$ stabilizes around
$0.03-0.05$, consistently with the amount of correction obtained
for $\sqrt s> 5\,TeV$.

\begin{figure}[b]
\begin{center}
\includegraphics[scale=0.55,angle=270]{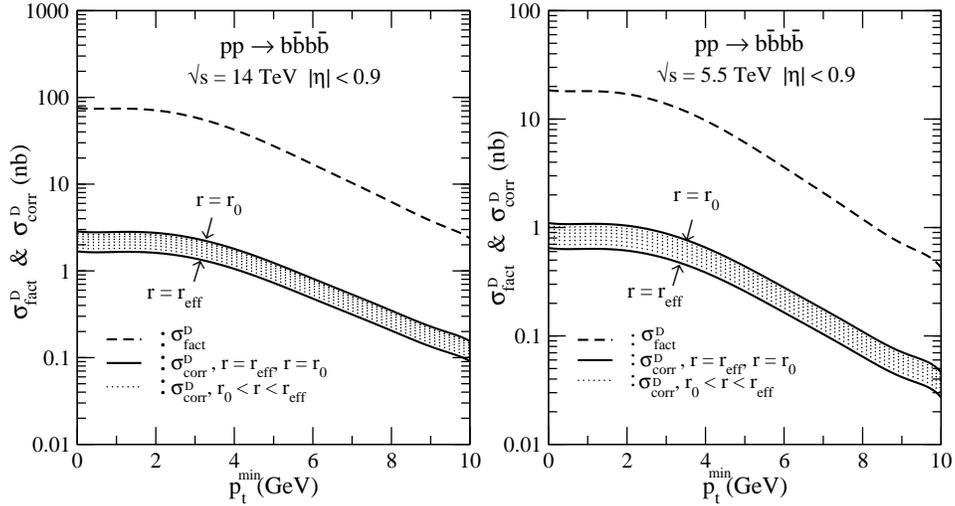}
\caption{$b\bar bb\bar b$ production cross-section at $\sqrt s=14\,TeV$ and $\sqrt
s=5.5\,TeV$ as a function of $p_t^{min}$, the minimum value of transverse momenta of the
outgoing $b$ quarks; the dashed curves refer to the double-parton scattering factorized
term $\sigma_{fact}^D$, while upper and lower solid curves refer to $\sigma_{corr}^D$
with geometrical factors obtained setting $r=r_0$ and $r=r_{eff}$ respectively.}
\label{bb-spectra-p0}
\end{center}
\end{figure}

\begin{figure}[t]
\begin{center}
\includegraphics[scale=0.55,angle=270]{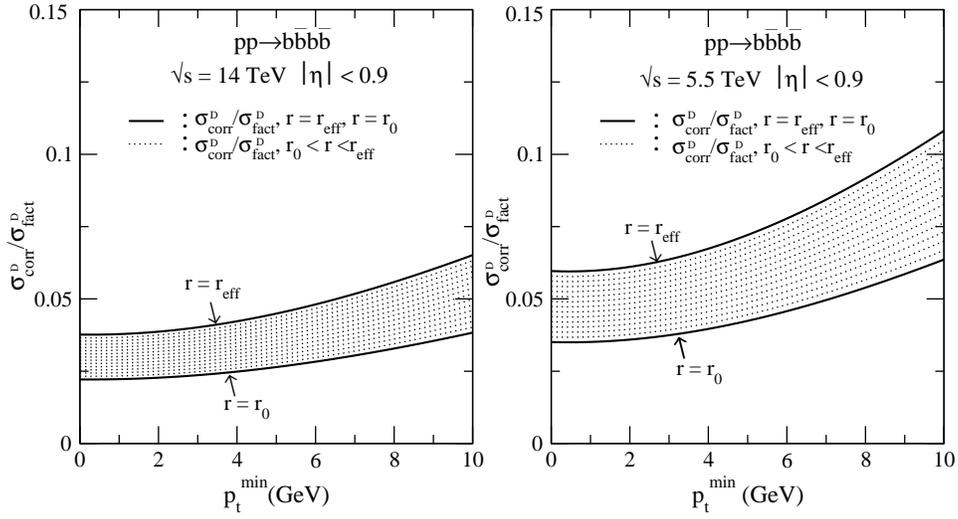}
\caption{ratio between $\sigma_{corr}^D$ and $\sigma_{fact}^D$ at $\sqrt s=14\,TeV$
(\emph{left-panel}) and $\sqrt s=5.5\,TeV$ (\emph{right-panel}) as a function of
$p_t^{min}$.}
\label{bb-p0-ratio}
\end{center}
\end{figure}

In Fig.[\ref{bb-spectra-p0}] we plot the $b\bar b b\bar b$
production cross-section at $\sqrt s=14\,TeV$ (\emph{left-panel})
and at $\sqrt s=5.5\,TeV$ (\emph{right-panel}), as a function of
the minimum value of transverse momenta of the outgoing $b$ quarks
$p_t^{min}$, in the pseudorapidity interval $|\eta|<0.9$. At
$\sqrt s = 14\,TeV$ with $p_t^{min}\in[0,10]\,GeV$ one has
$<x>\in[1.2 ,3.4 ]\times10^{-3}$, which leads to a contribution of
the correlation terms of the order of $(2-4)\%$ and of $(4-7)\%$,
respectively for the lower and the higher choices of $p_t^{min}$
Fig.(\ref{bb-p0-ratio})(\emph{left-panel}). At $\sqrt s =
5.5\,TeV$, in the considered range of variability of $p_t^{min}$
one has $<x>\in[3.5 ,6.5]\times10^{-3}$ and the contribution of
the correlation terms can become of the order of $12\%$,
Fig.(\ref{bb-p0-ratio})(\emph{right-panel}).

\begin{figure}[t]
\begin{center}
\includegraphics[scale=0.55,angle=270]{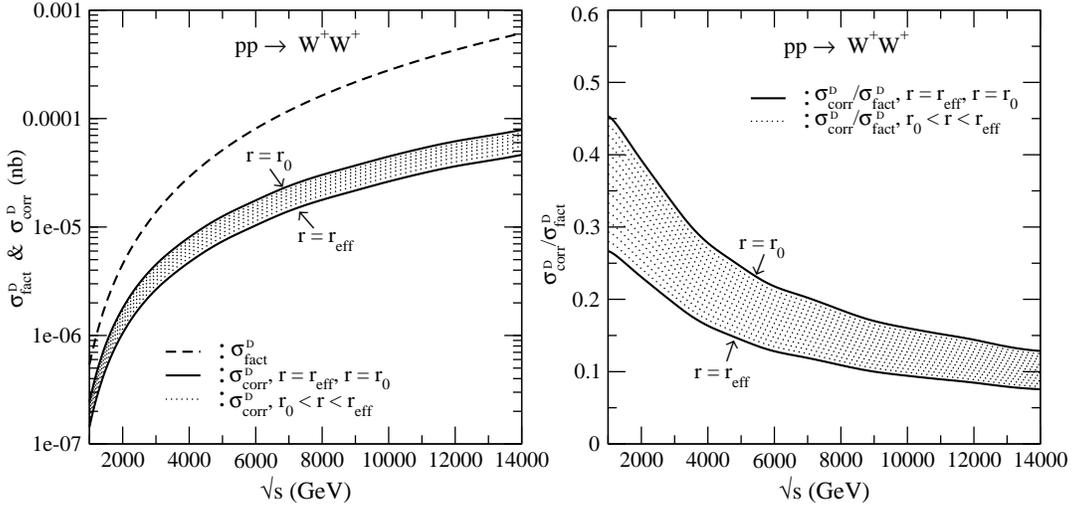}
\caption{\emph{Left-panel}: $W^+W^+$ inclusive cross-section as a function of the center
of mass energy; the dashed curve refers to the double-parton scattering factorized term
$\sigma_{fact}^D$, while upper and lower solid curves refer to $\sigma_{corr}^D$, with
geometrical factors obtained setting $r=r_0$ and $r=r_{eff}$ respectively. \\
\emph{Right-panel}: ratio between $\sigma_{corr}^D$ and $\sigma_{fact}^D$ as a function
of center of mass energy.}
\label{Wp-spectra}
\end{center}
\end{figure}

\begin{figure}[b]
\begin{center}
\includegraphics[scale=0.55,angle=270]{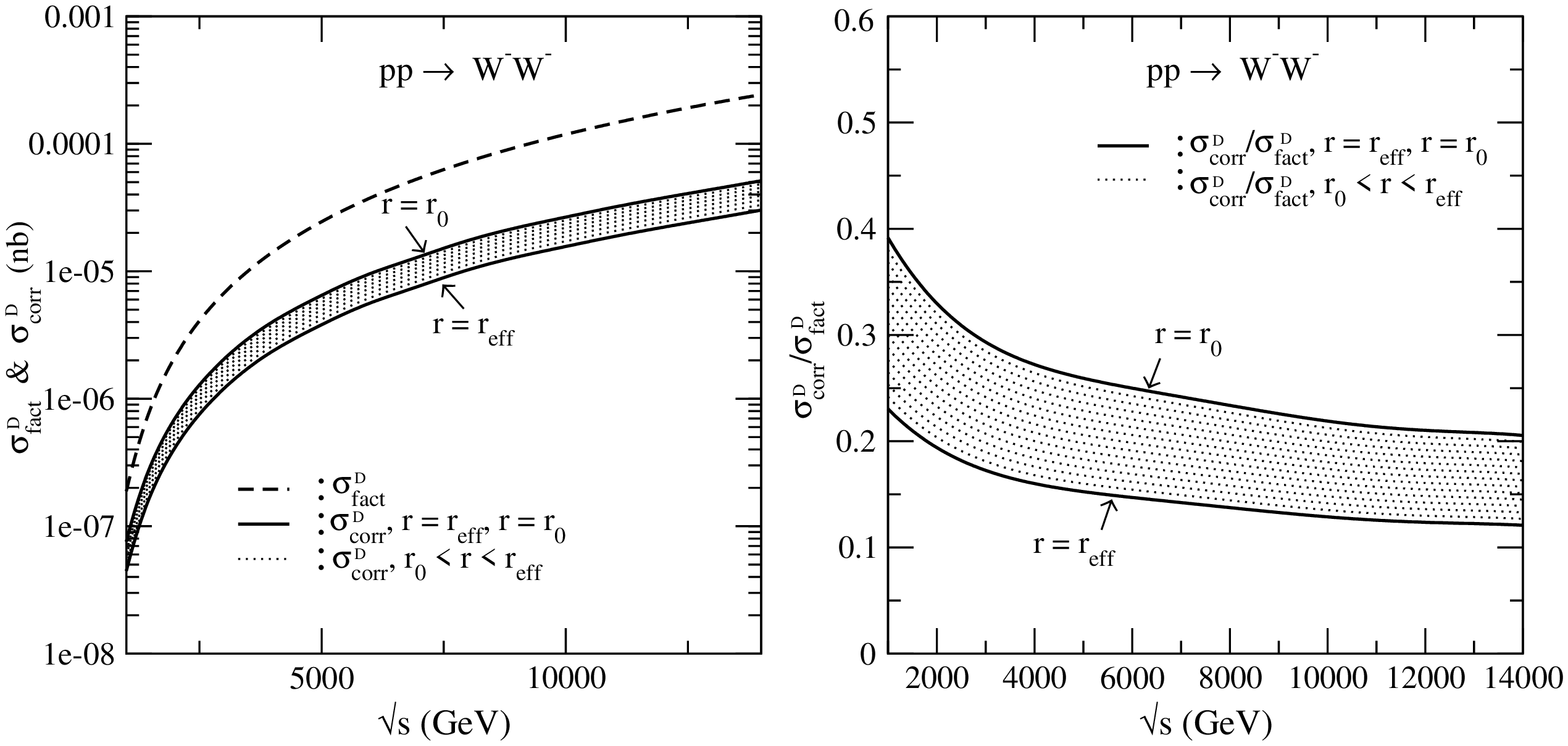}
\caption{\emph{Left-panel}: $W^-W^-$ inclusive cross-section as a function of the center
of mass energy; the dashed curve refers to the double-parton scattering factorized term
$\sigma_{fact}^D$, while the upper and the lower solid curves refer to the contribution
with correlations $\sigma_{corr}^D$, with geometrical factors obtained setting $r=r_0$
and $r=r_{eff}$ respectively. \\ \emph{Right-panel}: ratio between $\sigma_{corr}^D$ and
$\sigma_{fact}^D$ as a function of center of mass energy.}
\label{Wm-spectra}
\end{center}
\end{figure}

The cross sections of like-sign W pair production are evaluated at
the leading order, hence including only quark initiated processes
in the elementary interaction ($q\bar q'\to W$). Higher order
corrections are taken into account multiplying the lowest order
cross section by the factor $K\simeq
1+(8\,\pi/9)\,\alpha_s(M_W^2)$  \cite{Barger}. We plot in
fig.[\ref{Wp-spectra}] (\emph{left-panel}) the $W^+W^+$
cross-section as a function of the $pp$ center of mass energy. As
in the case of $b\bar b b\bar b$ production, the dashed curve
refers to the double-parton scattering factorized term
($\sigma_{fact}^D$), while the solid curves to the contribution of
the terms with correlations ($\sigma_{corr}^D$), for the two
different choices $r=r_0$ and $r=r_{eff}$. As one may infer from
the behavior of the qq-correlation ratio, for $<x>\in
[0.2,\,6]\times 10^{-2}$, which corresponds to the energy interval
considered, the corrections due to the correlation terms range
from $(27-45)\%$ at $\sqrt s=1\,TeV$ to $(7.5-13)\%$ at $\sqrt s =
14\,TeV$, depending on the choice of $\sigma_r$. The results for
$W^-W^-$ production are presented in fig.(\ref{Wm-spectra}). As
shown in the right panel the correlation terms can give
contributions ranging from $(23-40)\%$ at $1\,TeV$ to $(12-20)\%$
at $14\,TeV.$

\section{Conclusions}

As an effect of evolution, the multiparton distributions functions
are expected to become strongly correlated in momentum fraction at
large $Q^2$ and finite $x$ \cite{Kirschner,Shelest,Snigirev}. On
the other hand, the indications from the experimental observation
of multiparton collisions at Fermilab \cite{CDF} are not in favor
of strong correlation effects in fractional momenta. The most
likely reason being that the kinematical domain observed,
relatively low $x$ values and limited resolution scale, is far
from the limiting case considered in QCD.

The possibility of testing multiparton collisions at high
resolution scales at the LHC will open the opportunity of testing
the correlations predicted by evolution. To have an indication on
the importance of the effects to be expected, we have considered a
high resolution scale multiparton process (equal sign $W$ pair
production) and, for comparison, a sizably smaller resolution
scale process ($b\bar b b\bar b$ production) in $pp$ collisions in
the energy range $1{\rm TeV}\le\sqrt s\le 14{\rm Tev}$. In both
cases the production process may take place either by single
(connected) or by multiple (disconnected) hard parton collisions,
while the two contributions may be disentangled applying proper
cuts in the final state \cite{CDF, Stirling1, DelFabbro4}. To study
the effects of correlations we have hence worked out the
disconnected contributions to the cross sections after evolving
the multiparton distribution functions at high resolution scales.

Our result is that the contribution of the terms with
correlations, in equal sign $W$ pairs production, might be almost
40\% of the cross section at 1 TeV and might still be a 20\%
effect at the LHC. The effect is much smaller in $b\bar b b\bar b$
production, where corrections to the usually considered factorized
distribution are typically between 5 and 10\%.

\begin{acknowledgments}

This work was partially supported by the Italian Ministry of
University and of Scientific and Technological Researches (MIUR)
by the Grant COFIN2003.

\end{acknowledgments}


\newpage
\end{document}